\documentstyle[multicol,prl,aps,amsfonts,bm,epsfig]{revtex}

\newcommand{\bra}[1]{\mbox{$\langle{#1}|$}}
\newcommand{\ket}[1]{\mbox{$|{#1}\rangle$}}

\def\beq{\begin{equation}}
\def\eeq{\end{equation}}
\def\beqa{\begin{eqnarray}}
\def\eeqa{\end{eqnarray}}

\def\Tr{{\rm Tr}}

\newcounter{saveeqn}
\newcommand{\alpheqn}{\setcounter{saveeqn}{\value{equation}}%
\stepcounter{saveeqn}\setcounter{equation}{0}%
\renewcommand{\theequation}{\mbox{\arabic{saveeqn}\alph{equation}}}}
\newcommand{\reseteqn}{\setcounter{equation}{\value{saveeqn}}%
\renewcommand{\theequation}{\arabic{equation}}}
\def\beql{\alpheqn \beqa}
\def\eeql{\eeqa \reseteqn}

\begin{document}

\title{
Generalization of the entanglement of formation for multi-party systems
\thanks{Supported by the National Natural Science Foundation of China under Grant No.
69773052}}
\author{An Min WANG$^{1,2,3}$}
\address{CCAST(World Laboratory) P.O.Box 8730, Beijing 100080, People's Republic of China$^1$\\
and Laboratory of Quantum Communication and Quantum Computing\\
University of Science and Technology of China$^2$\\
Department of Modern Physics, University of Science and Technology of China\\
P.O. Box 4, Hefei 230027, People's Republic of China$^3$}
\maketitle

\begin{abstract}
{We present the generalization of the entanglement of formation for three-party systems in a pure state. For three qubit system we derive out its explicit and 
closed expression which is a linear combination of the binary entropy functions with various arguments, and these arguments are clearly determined in terms of 
the components of state vector of three qubits. As a reasonable measure of entanglement, the main behaviors and elementary properties of this generalized 
entanglement of formation are showed through discussing some important and interesting examples. Moreover, we propose how to extend our definition to a 
mixed state in according to the familiar idea. Then, we suggest the generalization of the entanglement of formation for multi-party systems which is consistent 
with the regular definition for two-party systems and our definition for three-party systems.}

\medskip
{\noindent}PACS: 03.65.Ud  03.67.-a  \vfill
\end{abstract}

\begin{multicols}{2}
Quantum entanglement is a subtle nonlocal correlation among the parts of a quantum system that has no classical analog \cite{Bennett1,Plenio1}. Thus 
entanglement is best characterized and quantified as a feature of the system that cannot be created through local operations that act on the different parts 
separately, or by means of classical communication among the parts. The entanglement of a two-part pure quantum state can be conveniently quantified by the 
number of Bell pairs that can be distilled by local operations and classical communication. At present, there are three most promising ideas for quantifying 
entanglement that is the entanglement of distillation \cite{Bennett1}, the entanglement of formation \cite{Bennett1,Bennett2,Wootters} and the relative entropy 
of entanglement \cite{Vedral1} mainly for two-party systems. But it is still not known completely if it is , at least, possible to express 
the entanglement of a pure state with more than two parts in terms of some such standard currency. In fact, a universal measure of many-particle and 
multi-party pure-state entanglement, if one can be formulated, would have many applications. The problem we address in this paper is just how to quantify 
entanglement of multi-party systems. 

First, our idea is from the knowledge that the entanglement of a quantum system is related with its correlation index of its subsystems. In fact, the relative 
entropy of entanglement for a quantum system of two-party is just defined in this way. For a three-party system, we have the correlation index between one 
party and the other two parties to be: $I_{X,Y\!Z}=S(\rho_X)+S(\rho_{Y\!Z})-S(\rho_{X\!Y\!Z})$, 
where $\rho_{X}$ and $\rho_{YZ}$ are respectively the reduced density matrices by partially tracing off any two-party, that is 
$\rho_{X}=\Tr_{Y\!Z}(\rho_{X\!Y\!Z})$, and by partially tracing off any one-party, that is $\rho_{Y\!Z}=\Tr_{X}(\rho_{X\!Y\!Z})$. $S(\rho_{X})\; 
(X\!=\!A,B,C)$ and $S(\rho_{Y\!Z})\;(Y\!Z\!=\!AB, BC, AC)$  are respectively von-Neumann entropies of 
$\rho_{X}$ and $\rho_{Y\!Z}$. To consider the correction among three-party, we have to add the internal entanglement between $Y$ and $Z$ party and 
average the result for rotating index of three parties. Because $\rho_{Y\!Z}$ is generally a mixed state, then we think (assume) the full correction index to be
\beqa
I_{A,B,C}&=&\frac{1}{3}\sum_{X}S(\rho_X)+\frac{1}{3}\sum_{YZ}S(\rho_{Y\!Z})+\frac{1}{3}\sum_{YZ}E(\rho_{Y\!Z})\nonumber\\
& &-S(\rho_{X\!Y\!Z})
\eeqa
where $E(\rho_{Y\!Z})$ is a entanglement measure between $Y$ and $Z$ party. Considering tow parts to contribute a factor $1/2$  for average,  and 
noticing von Neumann entropy of a pure state to be zero,  we have 

\noindent{\bf Definition 1:} For any pure tri-particle state (density matrix) $\rho_{ABC}^{\rm P}$, we define its generalization of the entanglement of 
formation is
\beqa
E_{GF}(\rho_{ABC}^{\rm P})&=&\frac{1}{6}\left[E_F(\rho_{AB})+S(\rho_{AB})+E_F(\rho_{AC})\right.\nonumber\\
& &\left. +S(\rho_{AC})+E_F(\rho_{BC})+S(\rho_{BC})\right]\nonumber\\
& &+\frac{1}{6}\left[S(\rho_A)+S(\rho_B)+S(\rho_C)\right] \label{GFfor3PP}
\eeqa
In the definition (\ref{GFfor3PP}), we have taken the entanglement of formation $E_F(\rho_{AB}),E_F(\rho_{AC})$ and $E_F(\rho_{BC})$ as 
entanglement measure  respectively for $\rho_{AB},\rho_{AC}$ and $\rho_{BC}$. If we take our modified relative entropy of entanglement, we can obtain 
the modified relative entropy of entanglement for tri-party systems \cite{My1}.  Moreover, we also account for physical meaning of the modified relative 
entropy of entanglement for multi-party systems from another physical idea in our paper. The consistency between two kind of physical ideas makes us sure its 
reasonableness. Just well known, the entanglement of formation is defined by
\beq
E_{F}(\rho_{X\!Y})=\min_{\{p_i,\rho[i]\}\in{\cal{D}}}\sum_{i} p_iS(\rho_{X}[i]) \label{EF}
\eeq
where the minimum in Eq.(\ref{EF}) is taken over the set ${\cal{D}}$ that includes all the possible decompositions of pure states of $\rho=\sum_i p_i\rho^i$. 
While von Neumann entropy for $\rho$ reads
\beq
S(\rho)=-\Tr(\rho\log\rho)=-\sum_i \lambda_i \log\lambda_i
\eeq
where $\lambda_i$ takes over all of (non-zero) eigenvalues of $\rho$. When $\rho_{X\!Y}$ is a pure state, $E_F(\rho_{X\!Y}^{\rm 
P})=S(\rho_{X})=S(\rho_{Y})$.

Many calculations in this paper are based on the following lemma:

\noindent{\bf Lemma 1:} For a two-qubit system in a pure state, its explicit expression of the entanglement of formation is 
\beq
E_F(\rho_{X\!Y}^{\rm P})=H\left(\frac{1-\xi_{X\!Y}}{2}\right)=H\left(\frac{1+\xi_{X\!Y}}{2}\right)
\eeq
where $H(x)$ is the binary entropy function
\beq
H(x)=-x\log_2 x-(1-x)\log_2(1-x)
\eeq
and $\xi_{X\!Y}$ is the norm of the polarization vectors $\bm{\xi}_X$ or $\bm{\xi}_Y$ of the reduced density matrix $\rho_X$ or $\rho_Y$. That is, we 
have used the fact that
\beq
\rho_{X,Y}=\Tr_{X,Y}(\rho_{X\!Y})=\frac{1}{2}(\sigma_0+\bm{\xi}_{X,Y}\cdot\bm{\sigma})
\eeq
for two qubits. Here $\bm{\sigma}$ is the Pauli spin matrix and $\sigma_0$ is $2\times 2$ identity matrix. When $\rho_{X\!Y}$ is a pure state density matrix, 
it is easy to verify 
$\bm{\xi}^2_X=\bm{\xi}^2_Y$ and so we denote $\xi_{X\!Y}=|\bm{\xi_X}|=|\bm{\xi}_Y|$. 
It is clear that $\xi_{X\!Y}$ is a kind of good concurrence of the entanglement of formation as it is a function of the density matrix. Its relation to the 
concurrence introduced by Wootters\cite{Wootters} is ${\cal{C}}^2=1-\xi_{X\!Y}^2$. This lemma had been proved in our paper \cite{My2}.

Now we can formulate the basic theorem of this paper.

\noindent{\bf Theorem 1:} For a tri-qubit system in a pure state, its explicit expression of the generalized entanglement of formation is 
\beqa
& &E_{GF}(\rho_{ABC})=\frac{1}{6}\left\{\sum_{i=1}^2\left[p_{AB}^{(i)}H\left(\frac{1-\xi_{AB}^{(i)}}{2}\right)\right.\right.\nonumber\\ 
& &\quad \left.+p_{AC}^{(i)}H\left(\frac{1-\xi_{AC}^{(i)}}{2}\right)+
p_{BC}^{(i)}H\left(\frac{1-\xi_{BC}^{(i)}}{2}\right)\right] \nonumber\\
& &\quad H\left(\lambda_{AB}[1]\right)+H\left(\lambda_{AC}[1]\right) +H\left(\lambda_{BC}[1]\right)\nonumber\\[10pt] 
& &\quad\left. +H\left(\frac{1-\xi_A}{2}\right)+H\left(\frac{1-\xi_B}{2}\right)+H\left(\frac{1-\xi_C}{2}\right)\right\}\label{GEE}
\eeqa
where 
\beql
p_{AB}^{(1)}=a{a^*} + c{c^*} + e{e^*} + g{g^*}\label{2RPDAB1}\\
p_{AC}^{(1)}=a{a^*} + b{b^*} + e{e^*} + f{f^*}\\
p_{BC}^{(1)}=a{a^*} + b{b^*} + c{c^*} + d{d^*}\label{2RPDBC1}
\eeql
\beq
p_{X\!Y}^{(2)}=1-p_{X\!Y}^{(1)}\quad (XY=AB,AC,BC)\label{2RPD2}
\eeq
\beql
\xi_{AB}^{(1)}&=&{\sqrt{1 - {4
         |ag-ce|^2}/{\left(p_{AB}^{(1)}\right)^2}}}\label{2RPVAB1}\\
\xi_{AB}^{(2)}&=&{\sqrt{1 - {4
         |bh-df|^2}/{\left(p_{AB}^{(2)}\right)^2}}}\label{2RPVAB2}\\
\xi_{AC}^{(1)}&=&{\sqrt{1 - {4
         |af-be|^2}/{\left(p_{AC}^{(1)}\right)^2}}}\label{2RPVAC1}\\
\xi_{AC}^{(2)}&=&{\sqrt{1 - {4
         |ch-dg|^2}/{\left(p_{AC}^{(2)}\right)^2}}}\label{2RPVAC2}\\
\xi_{BC}^{(1)}&=&{\sqrt{1 - {4
         |ad-bc|^2}/{\left(p_{BC}^{(1)}\right)^2}}}\label{2RPVBC1}\\
\xi_{BC}^{(2)}&=&{\sqrt{1 - {4
         |eh-fg|^2}/{\left(p_{BC}^{(2)}\right)^2}}}\label{2RPVBC2}   
\eeql
\beql
& &\lambda_{AB}[1]=
\frac{1}{2}\{1-[1-4(|ad\!-\!bc|^2+|af\!-\!ah|^2+|be\!-\!bg|^2
\nonumber\\
& &\quad+|cf-de|^2+|ch-dg|^2+|eh-fg|^2+(af-bg)\nonumber\\
& &\quad\times(a^*h^*-b^*e^*)+(a^*f^*-b^*g^*)(ah-be))]^{1/2}\}\label{2REVAB1}\\
& &\lambda_{AC}[1]= \frac{1}{2}\{1-[1-4(|ad\!-\!bc|^2+|ag\!-\!ce|^2+|ah\!-\!bg|^2\nonumber\\
& &\quad+|bh-df|^2+|cf-de|^2+|eh-fg|^2+(ah-de)\nonumber\\
& &\quad\times(b^*g^*-c^*f^*)+(a^*h^*-d^*e^*)(bg-cf))]^{1/2}\}\\
& &\lambda_{BC}[1]=\frac{1}{2}\{1-[1-4(|af\!-\!be|^2+|ag\!-\!ce|^2+|ah\!-\!bg|^2\nonumber\\  
& &\quad+|bh-df|^2+|cf-de|^2+|ch-dg|^2+(ah-cf)\nonumber\\
& &\quad\times(b^*g^*-d^*e^*)+(a^*h^*-c^*f^*)(bg-de))]^{1/2}\}\label{2REVBC1}
\eeql
\beql
& &\xi_A=\{1\! - \!4\left[\left( aa^*\! +\! bb^*\! +\! cc^*\! +\! dd^*\!\right) 
          \left(ee^*\!+\!ff^*\! +\! gg^*\! +\! hh^*\!\right)\right.\nonumber\\
    & &\left.-\left( e{a^*}\! +\! f{b^*}\! +\! g{c^*}\! +\! h{d^*}\!\right) \left( a{e^*}\! +\! b{f^*}\! +\!c{g^*}\! +\! d{h^*}\! \right) \right]\}^{1/2} 
\label{NPVAof1R}\\
& &\xi_B=\{1\! -\! 4\left[\left(\! a{a^*}\! +\! b{b^*}\! +\! e{e^*}\! +\! 
            f{f^*}\!\right) 
          \left( c{c^*}\! +\! d{d^*}\! +\! g{g^*}\! +\! 
            h{h^*}\! \right)\right.\nonumber\\
& &\left.-\left( c{a^*}\! +\! 
              d{b^*}\! +\! g{e^*}\! +\! h{f^*}\! \right) 
            \left( a{c^*}\! +\! b{d^*}\! +\! e{g^*} + 
              f{h^*}\! \right)\right]\}^{1/2} \label{NPVBof1R}\\
& &\xi_C= \{1\! -\! 4\left[\left( a{a^*}\! +\! c{c^*}\! +\! e{e^*}\! +\! 
            g{g^*}\! \right) 
          \left( b{b^*}\! +\! d{d^*}\! +\! f{f^*}\! +\! 
            h{h^*}\! \right)\right.\nonumber\\ 
& &\left.-\left( b{a^*}\! +\! 
              d{c^*} \!+\! f{e^*}\! +\! h{g^*}\! \right) 
            \left( a{b^*}\! +\! c{d^*}\! +\! e{f^*}\! +\! 
              g{h^*}\! \right) \right]\}^{1/2} \label{NPVCof1R}
\eeql
Note that if some $p_{X\!Y}^{(i)}, (XY=AB,AC,BC;\,i=1,2)$ are zero, then the corresponding $\xi_{X\!Y}^{(i)}$ are also zero since 
$p_{X\!Y}^{(i)}\rho_{X\!Y}^{(i)}$ are zero matrices. In addition, we always take  $\displaystyle\lim_{x->0}x\log x=0$. 

\noindent{\bf Proof:} Without loss of generality, we denote the pure state for a tri-qubit system as
\beqa
\ket{\psi_{ABC}}&=&a\ket{000}+b\ket{001}+c\ket{010}+d\ket{011}\nonumber\\
& &+e\ket{100}+f\ket{101}+g\ket{110}+h\ket{111}
\eeqa
Its density matrix is then $\ket{\psi_{ABC}}\bra{\psi_{ABC}}$. We can find the reduced density matrices tracing off one-party and obtain their pure state 
decompositions
\beq
\rho_{X\!Y}=p_{X\!Y}^{(1)}\rho_{X\!Y}^{(1)}+p_{X\!Y}^{(2)}\rho_{X\!Y}^{(2)}\eeq
where $p_{X\!Y}^{(i)},(XY=AB,AC,BC;
i=1,2)$ are given in Eq.(\ref{2RPDAB1}-\ref{2RPDBC1},\ref{2RPD2}) and 
\beql
\rho_{AB}^{(1)}&=&\frac{1}{p_{AB}^{(1)}}\left(a\ket{00}+c\ket{01}+e\ket{10}+g\ket{11}\right)\nonumber\\& 
&\left(a^*\bra{00}+c^*\bra{01}+e^*\bra{10}+g^*\bra{11}\right)\\
\rho_{AB}^{(2)}&=&\frac{1}{p_{AB}^{(2)}}\left(b\ket{00}+d\ket{01}+f\ket{10}+h\ket{11}\right)\nonumber\\& 
&\left(b^*\bra{00}+d^*\bra{01}+f^*\bra{10}+h^*\bra{11}\right)\\
\rho_{AC}^{(1)}&=&\frac{1}{p_{AC}^{(1)}}\left(a\ket{00}+b\ket{01}+e\ket{10}+f\ket{11}\right)\nonumber\\& 
&\left(a^*\bra{00}+b^*\bra{01}+e^*\bra{10}+f^*\bra{11}\right)\\
\rho_{AC}^{(2)}&=&\frac{1}{p_{AC}^{(2)}}\left(c\ket{00}+d\ket{01}+g\ket{10}+h\ket{11}\right)\nonumber\\& 
&\left(c^*\bra{00}+d^*\bra{01}+g^*\bra{10}+h^*\bra{11}\right)\\
\rho_{BC}^{(1)}&=&\frac{1}{p_{BC}^{(1)}}\left(a\ket{00}+b\ket{01}+c\ket{10}+d\ket{11}\right)\nonumber\\& 
&\left(a^*\bra{00}+b^*\bra{01}+c^*\bra{10}+d^*\bra{11}\right)\\
\rho_{BC}^{(2)}&=&\frac{1}{p_{BC}^{(2)}}\left(e\ket{00}+f\ket{01}+g\ket{10}+h\ket{11}\right)\nonumber\\ & 
&\left(e^*\bra{00}+f^*\bra{01}+g^*\bra{10}+h^*\bra{11}\right)
\eeql
Obviously, if some $p_{X\!Y}^{(i)}, (XY=AB,AC,BC;\,i=1\ {\rm or}\ 2)$ are zero, then the corresponding $\rho_{X\!Y}$ are just pure states. But for a 
fixed $XY$, $p_{X\!Y}^{(1)}$ and $p_{X\!Y}^{(2)}$ can not be zero at the same time, otherwise $\rho_{ABC}$ is a zero matrix without any significance. 
We guess these pure state decompositions corresponding to one which can minimize the quantity $\sum_i p_{X\!Y}^{(i)}S(\rho_{X}^{(i)})$ (unproved). In 
fact, it is indeed so at least for most interesting systems. It must be emphasized that all of the reduced density matrices $\rho_{AB},\rho_{AC},\rho_{BC}$ at 
most have two non-zero eigenvalues. One of eigenvalues of them are given out in Eq.(\ref{2REVAB1}-\ref{2REVBC1}). Another eigenvalues of them are 
equal to $1-\lambda_{X\!Y}[1]$ \cite{My3}. 

In the same way, we can find the reduced density matrices tracing off two-party 
and obtain all the polarized vectors for them. The useful quantities are the norm of the polarized vectors in Eqs. (\ref{NPVAof1R}-\ref{NPVCof1R}). 

In terms of the definition of entanglement of formation, for $XY=AB,AC,BC$, we have 
\beqa
& &E_F(\rho_{X\!Y})=\sum_{i=1}^2 p_{X\!Y}^{(i)}E_F(\rho_{X\!Y}^{(i)})=\sum_{i=1}^2 p_{X\!Y}^{(i)} 
S\left[\Tr_X(\rho_{X\!Y}^{(i)})\right]\nonumber\\
& &\quad =\sum_{i=1}^2 p_{X\!Y}^{(i)} H\left(\frac{1-\xi_{X\!Y}^{(i)}}{2}\right)=\sum_{i=1}^2 
p_{X\!Y}^{(i)}H\left(\frac{1+\xi_{X\!Y}^{(i)}}{2}\right)
\eeqa
From $\rho_{X\!Y}$ at most has two non-zero eigenvalues and their summation is 1, it follows that
\beq
S(\rho_{X\!Y})=H\left(\lambda_{X\!Y}[1]\right)=H\left(\lambda_{X\!Y}[2]\right)
\eeq
Because the eigenvalue of $2\times 2$ reduced density matrices $\rho_A,\rho_B$ and $\rho_C$ are respectively $(1\pm\xi_{A})/2,(1\pm\xi_{B})/2$ and 
$(1\pm\xi_{C})/2$, we know their von Neumann entropies are
\beq
S(\rho_{X})=H\left(\frac{1-\xi_{X}}{2}\right)\quad (X=A,B,C)
\eeq

Combining all the results as stated above into the definition of the generalized entanglement of formation, we can finish the proof of the theorem one.

Thus, it is important and interesting whether our definition of the generalized entanglement of formation is reasonable. Obviously, based on the fact that the 
maximum value of the binary entropy function is 1 when its argument between 0 and 1, the theorem one results in that the maximum value of the generalized 
entanglement of formation for the three-qubit system seems to be $3/2$. But we think that its maximum value is actually is 1 since the various arguments in these 
binary entropy functions are related. The strictly proof will expect to be obtained in future. In fact, we find this maximum value is saturated by all of four pairs of 
GHZ ``cat" states $
\ket{\phi^{\rm GHZ}}(\pm)=\{(\ket{000}\pm\ket{111})/\sqrt{2}$,
$(\ket{001}\pm\ket{110})/\sqrt{2}$,$\ket{010}\pm\ket{101})/\sqrt{2}$,
$(\ket{011}\pm\ket{100})/\sqrt{2}\}$. For 
12 kinds of the extended Bell states $
\ket{\psi^{\rm EB}_{AB}}=\ket{\phi^{\rm B}_{AB}}\otimes\ket{\chi_C}$, $\ket{\psi^{\rm 
EB}_{AC\;1}}=(\ket{0}_A\ket{\chi_B}\ket{0}_C\pm\ket{1}_A\ket{\chi_B}\ket{1}_C)/\sqrt{2}$,$\ket{\psi^{\rm 
EB}_{AC\;2}}=(\ket{0}_A\ket{\chi_B}\ket{1}_C\pm\ket{1}_A\ket{\chi_B}\ket{0}_C)/\sqrt{2}$,$\ket{\psi^{\rm 
EB}_{BC}}=\ket{\chi_A}\otimes\ket{\phi^{\rm B}_{BC}}$, the generalized entanglement of formation are 5/6. While for all the separable states, we can 
find that their generalized entanglement of formation are zero, that is equal to its minimum values. This is because for a tri-party system in pure state, if it is 
separable, it can be written as $\rho_A\otimes\rho_B\otimes\rho_C$. This means that all the 6 reduced matrices are the pure states and so their von Neumann 
entropies are equal to zero. Moreover, all 3 the reduced density matrices tracing off one-party are also separable, their entanglement of formation vanish. 
Therefore the generalized entanglement of formation for a separable state must be nothing. In general cases, we can seen the generalized entanglement of 
formation has indeed good behaviors. For example, for the GHZ-like state $a\ket{000}+h\ket{111}$, the generalized entanglement of formation is equal to the 
binary entropy function $H(a a^*)=H(h h^*)$. This is just what we expect.   
For the state
\beqa
\ket{\psi_{ABC}}&=&\frac{x}{3}\ket{000}+\frac{\sqrt{2-x^2}}{3}\ket{001}+\frac{1}{3}\ket{010}\nonumber\\
& &+\frac{1}{\sqrt{6}}\ket{101}+\frac{1}{\sqrt{6}}\ket{110}+\frac{1}{\sqrt{3}}\ket{111}
\eeqa
when $x$ varies from 0 to $\sqrt{2}$, its generalized entanglement of formation is displayed in the following figure:
\begin{center}
\epsfxsize=3.0in
\epsfysize=1.8in
\epsfbox{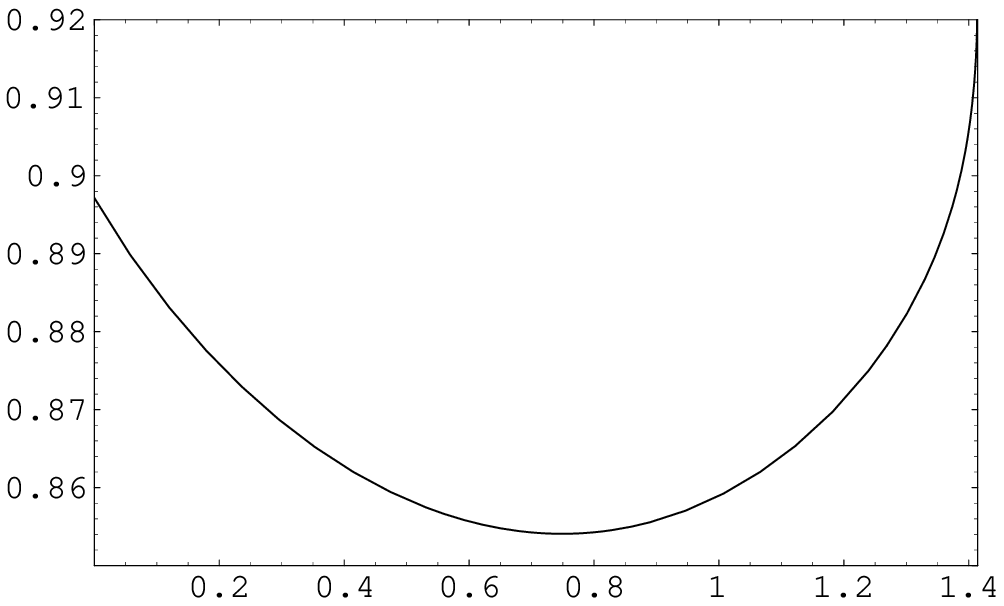}
\end{center}
Of course, it can prove $E_{GF}$'s invariance under local unitary operations from its definition and, for the system made of qubits, monotonicity under local 
operations and  classical communication from its explicit expression (\ref{GEE}) \cite{My2} because it is a linear combination of binary entropy functions. 

Note the fact that the reduced density matrices for a multi-party system are in general can be the mixed states, just like seen in above tri-party systems, we have 
to know how to define the generalized entanglement of formation for the mixed states. According to Bennett $et.\; al$'s idea \cite{Bennett1,Bennett2}, we can 
propose:

\noindent{\bf Definition 2:} For a three-party systems in the mixed state with a serial of pure state decompositions
\beq
\rho_{ABC}^{\rm M}=\sum_i p_i \rho_{ABC}^{\rm P}[i] \label{GFfor3PM}
\eeq
which form a set ${\cal{D}}$. 
Its generalized entanglement of formation is defined by
\beq
E_{GF}(\rho_{ABC}^{\rm M})=\min_{\{p_i,\rho[i]\}\in{\cal{D}}}\sum_{i} p_i E_{GF}(\rho_{ABC}^{\rm P}[i])
\eeq
The condition to take the minimum is necessary since there exist such some states that the different decompositions have the different values of the generalized 
entanglement of formation, for example, the state  
$\frac{1}{2}(\ket{000}\bra{000}+\ket{111}\bra{111})=\frac{1}{4}(\ket{000}+\ket{111})(\bra{000}+\bra{111})+\frac{1}{4}(\ket{000}-\ket{111})(\
\bra{000}-\bra{111})$. 

Now, we can further suggest (conjecture) the generalized entanglement of formation for multi-party systems. Because there are $2^{n-1}-1$ ways to 
decompose a compound system $A_1A_2\cdots A_n$ into its two parts (subsystems). For each part, if it is not a simple party, we have to add the internal 
entanglement. Considering two parts to contribute a factor $1/2$  for average,  and noticing von Neumann entropy of a pure state to be zero,  we have

\noindent{\bf Definition 3:} For a $n$-party system in a pure state, its generalized entanglement of formation can be defined by
\beqa
& &E_{GF}(\rho_{A_1A_2\cdots A_n})=\frac{1}{2^n-2}\sum_{m=1}^{n-1}\sum_{i_1,i_2,\cdots,i_m=1\atop i_1<i_2<\cdots<i_m}^{n}\nonumber\\ & 
& \qquad
\left\{E_{GF}[\Tr_{A_{i_1}A_{i_2}\cdots A_{i_m}}(\rho_{A_1A_2\cdots A_n})]\right.\nonumber\\
& &\qquad\left.+S[\Tr_{A_{i_1}A_{i_2}\cdots A_{i_m}}(\rho_{A_1A_2\cdots A_n})]\right\}
\eeqa 
Here, we have set that $E_{GF}\left(\Tr_{A_{i_1}A_{i_2}\cdots A_{i_n-1}}(\rho_{A_1A_2\cdots A_n})\right)=E(\rho_{A_j})=0$ $(j\neq 
i_1,i_2,\cdots,i_{n-1})$ because of no any entanglement for one-party systems. And, the factor $2^n-2$ comes from the number of all of the reduced density 
matrices. For a $n$-party in a mixed state,  again by use of Bennett $et.\; al$'s idea we define 
\beq
E_{GF}(\rho_{A_1\!A_2\!\cdots\! A_n}^{\rm M})=\min_{\{p_i,\!\rho[i]\}\!\in\!{\cal{D}}}\sum_{i} p_i E_{GF}(\rho_{A_1\!A_2\!\cdots\! A_n}^{\rm 
P}\![i])
\eeq
where ${\cal{D}}$ is a set of all possible pure state decompositions. It is easy to see, for two-party system this definition is just regular definition (\ref{EF}) 
since $S(\Tr_A(\rho_{AB}[i]))=S(\Tr_B(\rho_{AB}[i]))$ when $\rho_{AB}[i]$ is a pure state. For a three-party system, we also back to our definition 
(\ref{GFfor3PP}) or (\ref{GFfor3PM}).

In the above definition, we can see that the generalized entanglement of formation for a $n$-party system is given by an equal weight $1/(2^n-2)$ linear 
combination by  $2^n-2-n$ generalized entanglement of formation for from two-party to $(n-1)$-party systems as well $2^n-2$ von Neumann entropies for 
from one-party to $(n-1)$-party systems, in which, for the $i$-party the corresponding number is $n!/[i!(n-i)!]$ which is equal to the number of the reduced 
matrices for the $i$-party. In calculation, we need to reduce all generalized entanglement of formation for multi-party systems up to one for two-party systems 
step by step. The last, we can obtain an expression only depending on a linear combination of many von Neumann entropies. Of course, the final value of the 
generalized entanglement of formation can be given in principle. However, during this process, we will encounter the very difficult problem to find the minimum 
pure state decomposition. At present, no an effective method appears. Consequently, it is absolutely not a simple task to calculate the generalized entanglement 
of formation for multi-party more than three-party systems even if we have above definitions. Moreover, we do not consider the case to decompose a 
compound system into more parts than two based the experience from tri-party systems.  Actually, our modified relative entropy of entanglement can solve 
these difficulties in principle \cite{My1} which can be seen by the interesting readers.

\end{multicols}

\end{document}